# Arago spot formation in the time domain

**C Finot [1] and H Rigneault [2]**
[1] Laboratoire Interdisciplinaire Carnot de Bourgogne, UMR 6303 CNRS - Université de Bourgogne-Franche-Comté, 9 avenue Alain Savary, BP 47870, 21078 Dijon Cedex, France
[2] Aix Marseille Université, CNRS, Centrale Marseille, Institut Fresnel UMR 7249, 13397 Marseille, France

E-mail : christophe.finot@u-bourgogne.fr

**Abstract.** Using the space-time duality we theoretically and experimentally revisit the Arago spot formation in the time domain and explore the temporal emergence of a bright spot in the time shadow of an initial waveform. We describe the linear regime of propagation using Fresnel's integrals and show that in the presence of Kerr nonlinearity the Arago spot formation is affected depending on the sign of the dispersion. We finally confirm experimentally the Arago spot formation in the time domain using a telecommunication optical fiber platform.

## 1. Introduction

The history of conceptions of light is among the most exciting scientific adventures. Debates on the corpuscular or wave nature of light was thus a subject of crucial importance for several centuries, until its dual nature was fully understood. If Young's experiments led to significant progress in understanding the wave behavior of light [1, 2], it was Fresnel's work that made it possible to establish the most solid theoretical basis of wave optics [3]. The discussions at the French 'Académie des Sciences' held two centuries ago remained particularly famous concerning the experiment that validated Fresnel's approach: according to Fresnel's diffraction theory a counter-intuitive bright spot should appear at the center of the geometric shadow of an illuminated opaque circular object [4]. This phenomenon has remained known as the Arago or Poisson's spot, named after the members of the jury that discussed the results. Since then, Fresnel's wave theory has been widely accepted to describe wave propagation in space and time. Indeed, dispersion and one-dimensional diffraction are linked by a similar mathematical formalism, so that a powerful analogy exists between the temporal and spatial propagation of waves [5-8]. Temporal analogues of common optical systems have been proposed such as lenses [9], imaging systems [6, 10], diffraction gratings [11], lenticular lenses [12], interferometric devices [13] or Talbot effect [14]. This opens up a whole range of new possibilities for ultrafast photonics.

In this paper, we revisit the Arago spot formation in the time domain through the temporal evolution of light after being briefly stopped by an obstacle. We first recall the spatial/temporal analogy and the modelling of the temporal propagation of light in an optical waveguide with possible Kerr non-linearity. We then discuss the results of numerical simulations for a time domain Arago experiment and confirm the emergence of light where it was initially absent. As the power increases, we show that the Arago spot intensity is affected by the sign of the dispersion. We finally confirm experimentally the existence of the Arago spot formation in the time domain using telecom optical fibers and fast optoelectronics.

## 2. Situation under investigation and emergence of a bright spot
### 2.1 Time-space duality in linear optics and propagation

Before discussing our experiments, let us first recall the basis of the analogy between the spatial evolution of light affected by diffraction and the temporal changes experienced by light when dispersion is involved. We consider the simple case where a monochromatic plane wave with wavelength $\lambda$ and an amplitude $a_0$ illuminates an opaque screen with a width $2l$ and an infinite length. In this 1D transverse problem that is illustrated on Fig. 1(a), the longitudinal evolution of light $a(x,z)$ in the scalar approximation is ruled by the following differential equation :

$$i\frac{\partial a}{\partial z} = -\frac{1}{2k_0}\frac{\partial^2 a}{\partial x^2}, \tag{1}$$

with $x$ and $z$ being the transverse and longitudinal coordinates respectively and $k_0 = 2\pi/\lambda$ the wavenumber. The two edges of the opaque screen diffract the light leading to the progressive emergence of a bright spot at the center of the screen shadow. Note that this spot is less intense and spatially broader that in the historic Arago experiment conducted in the space domain and using a 2D opaque circular screen. Indeed in 2D, the Arago bright spot results as a constructive interference from the scattered light at the edge of the 2D opaque screen.

The goal of the present paper is to study the temporal equivalent of a 1D Arago experiment where the stop is an opaque slit. We therefore consider a continuous wave where light has been switched off for a duration $2\,T_0$ as illustrated in Fig. 1(b1) and modelled by the function $a(t,z=0) = a_0\,[1\text{-rect}(t/2T_0)]$ where rect is the rectangular function of width 1 and $t$ being the temporal coordinate. This shaped temporal waveform then propagates in a dispersive single mode waveguide, typically an optical fiber, that ensures that its spatial transverse profile is unaffected upon propagation. The temporal profile of the light in the approximation of the slowly varying envelope evolves according to:

$$i\frac{\partial a}{\partial z} = \frac{1}{2}\beta_2\frac{\partial^2 a}{\partial t^2}, \tag{2}$$

with $\beta_2$ the second order dispersion coefficient. After propagation in the waveguide, dispersion leads to the alteration of the temporal profile (Fig. 1(b2), blue curve): the initially very sharp edges of the waveform have been smoothened, strong oscillations have appeared and a light intensity increase has emerged at the dark pulse center where initially no light was present. This last feature being the temporal analog of the Arago spot in this 1D experiment.

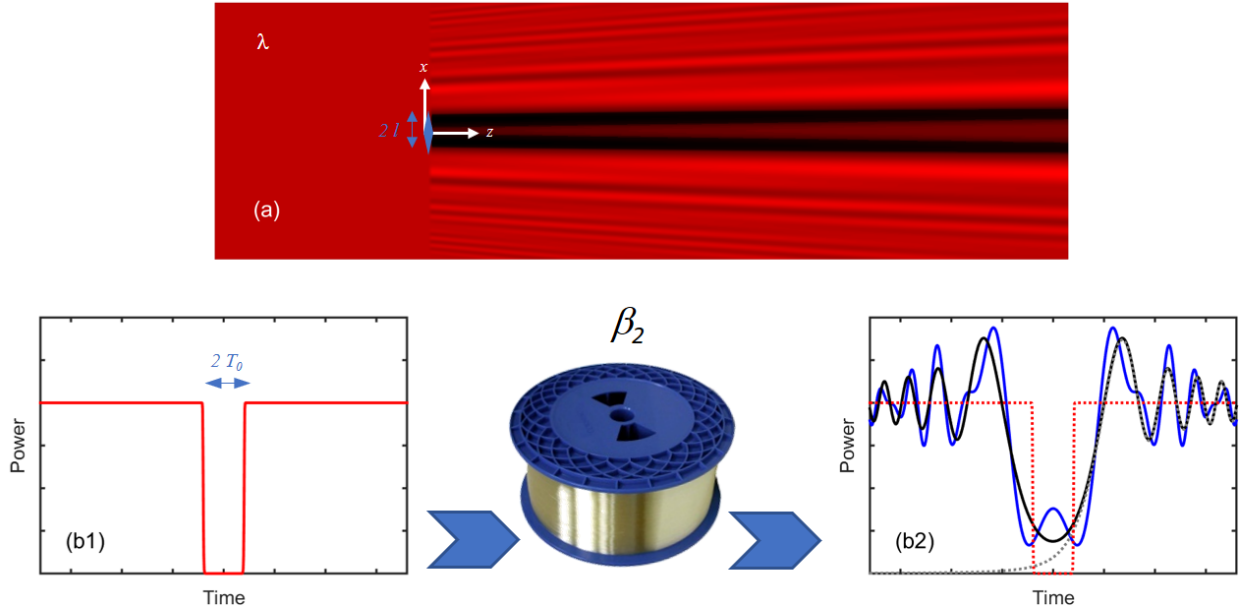

**Figure 1** – (a) Illustration of the longitudinal evolution of the diffraction pattern of a monochromatic wave affected by the presence by an opaque screen of width $2l$. (b) Evolution of the temporal intensity profile of a hole of light. The initial profile (b1) is affected by dispersion. The resulting pattern (b2, blue line) is compared with the shape of diffraction by a semi-infinite screen (grey dotted line) and with the superposition of the intensity pattern resulting from the two edges (solid black line).

The space-time duality readily appears in the mathematical structure of equations (1) and (2) where space and time are exchanged and the link between diffraction and dispersion is clear [15]. Indeed, both effects fulfill the same normalized differential equation:

$$i\frac{\partial \psi}{\partial \xi} = -\frac{1}{2}\frac{\partial^2 \psi}{\partial \eta^2}, \qquad (3)$$

where $\xi$ is a normalized propagation distance $\xi = z / L_D$ where $L_D$ is defined by $L_D = -T_0^2/\beta_2$ (for dispersion, note that with this convention $L_D$ is negative for normally dispersive fibers and positive for anomalous dispersion) or $L_D = k_0 l^2$ (for diffraction). $\eta$ is the normalized time $\eta = t/T_0$ (dispersion) or the normalized transverse coordinate $\eta = x/l$ (diffraction). $\psi = a / a_0$ is the field $a$ normalized with respect to the amplitude of the initial plane wave. Consequently, both diffraction and dispersion implies the development of a quadratic spectral phase and lead formally to similar consequences. This space-time duality is useful to better understand the temporal evolution of temporally sculpted short pulses subject to dispersion. As a first rough approximation, we can consider that the pattern created by the opaque temporal zone is the result of the superposition of the patterns created by a semi-infinite edges located at $\eta = \pm 1$. The diffraction pattern of a semi-infinite screen is indeed among the first examples that are taught to students when introducing diffraction [16, 17]. For this case, Fraunhofer approximations do not hold but the edge problem can be solved analytically by involving Fresnel's integrals and the graphical Cornu's spiral plot. Indeed, the field $\psi_E$ diffracted by a single initial sharp edge centered at $\eta = 0$ is given by :

$$\psi_E(u) = 0.5 + C(u) + i\left(0.5 + S(u)\right) \qquad (4)$$

where $u = \eta / (\pi \xi)^{1/2}$ and $C$ and $S$ are the Fresnel integrals. This semi-infinite screen (grey dashed line in Fig. 1(b1)) brings some qualitative insight to understand the softening of the edges as well as the nature of the strong fluctuations that develop on each side of the edge in the blue curve of Fig. 1(b1). However, the addition of the temporal intensity profiles created by each intensity jump (black line in Fig. 1(b1)) cannot account for the development of the central spot that is the result of a constructive interference process: given the symmetry of our problem, the patterns created by each edge are in phase at the center of the waveform, leading to an increase by a factor 2 of the intensity profile. This intensity increase is the 1D temporal Arago bright spot.

## 2.2 Evolution in a purely dispersive medium

The field diffracted by the opaque 1D screen can be fully analytically expressed by means of Fresnel integrals:

$$\psi_S(\eta,\xi) = \frac{1}{\sqrt{2}}\left[\left(1 + C\left(\frac{\eta-1}{\sqrt{\pi\xi}}\right) + C\left(\frac{-\eta-1}{\sqrt{\pi\xi}}\right)\right) + i\left(1 + S\left(\frac{\eta-1}{\sqrt{\pi\xi}}\right) + S\left(\frac{-\eta-1}{\sqrt{\pi\xi}}\right)\right)\right] \quad (5)$$

We consider here the time domain evolution and we have plotted on Fig. 2(a) the temporal amplitude and phase profiles of the wave at three different propagation distances for an anomalous regime of propagation (positive $L_D$). We can make out that the amplitude of the ripple observed on the top level of the wave does not evolve much with propagation distance. The temporal position of the maximum of the oscillations tend to move away from the center of the waveform. Those trends are consistent with the diffraction pattern of a straight edge (dash line in Fig. 2(b1)) and we can note that the maximum value of the oscillation is obtained for $\eta = \pm(1 + 2.12\,\xi^{1/2})$ with a value of 1.37 expected in this case [18]. We concentrate now on the Arago spot formation located at the center of the diffracted pattern where one can notice the continuous increase of a central spot with increasing propagation distance (Fig 2(a1)). The growth of the central spot is better seen on the longitudinal intensity evolution map plotted on Fig. 2(b1) and it can be quite easily expressed analytically. Using $Z = (\pi\,\xi)^{-1/2}$, the field at the center of the waveform is

$$\psi_{S,0}(Z) = \sqrt{2}\left[(0.5 - C(Z)) + i(0.5 - S(Z))\right] \quad (6)$$

or equivalently, using the complex error function erf:

$$\psi_{S,0}(Z) = \frac{1+i}{\sqrt{2}}\left[1 - erf\left(\frac{1-i}{2}\sqrt{\pi}\,Z\right)\right] \quad (7)$$

The intensity profiles at the center of the waveform then evolves as:

$$|\psi_{S,0}(Z)|^2 = 2\left[(C(Z) - 0.5)^2 + (S(Z) - 0.5)^2\right] \quad (8)$$

and can be graphically predicted using the Cornu's spiral as directly linked to the distance between the point of curvilinear coordinate $Z$ and the fixed point $(1/2, 1/2)$. The intensity profile at the center of the waveform therefore continuously increases up to a value that becomes equals to the intensity outside of shadow.

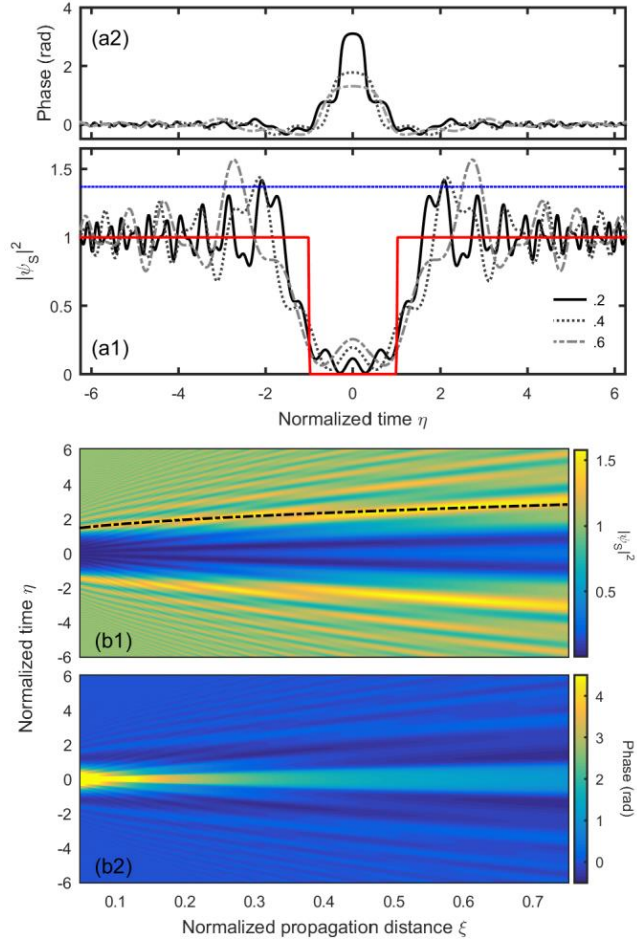

**Figure 2** – (a) Temporal intensity and phase profiles at three propagation distances $\xi$ =0.2, 0.4 and 0.6. Results are compared with the initial intensity profile (red curve). (b) Longitudinal evolution of the temporal intensity and phase profiles (b1 and b2, respectively). The black dashed line is the analytical prediction of the first maximum of the pattern for a single edge diffraction. Results are here obtained with anomalous dispersion (positive $L_D$) and are based on the analytical solution provided by Eq. (5).

We complement this study by plotting the temporal phase profile of the diffracted field (panels (a2) and (b2) of Fig. 2). Similarly to the diffraction of a straight edge [18], we can notice a significant phase difference between the central part and the plateau. Whereas the intensity profile is insensitive to the sign of the dispersion (normal or anomalous), the sign of the phase offset $\Delta\varphi$ that exists between the central part and the plateau is influenced by the sign of the dispersion. Similarly, to the intensity profile, the evolution of the phase offset can be expressed analytically as:

$$\Delta\varphi = \arctan\left(\frac{1-2\,S(Z)}{1-2\,C(Z)}\right) - \frac{\pi}{4} \qquad (9)$$

We can note that the phase difference $|\Delta\varphi|$ tends to decrease asymptotically towards 0.

### 2.3 Evolution with nonlinearity

Contrary to the usual diffraction in free space, propagation in a waveguide can also involve nonlinear effect. Indeed, the temporal evolution of a waveform in an optical fiber is affected by Kerr nonlinearity that can be taken into account through an additional term accounting for self-phase modulation in Eq. (2), leading to the well-known nonlinear Schrödinger equation (NLSE) [19]:

$$i\frac{\partial a}{\partial z} = \frac{1}{2}\beta_2 \frac{\partial^2 a}{\partial t^2} - \gamma |a|^2 a, \tag{10}$$

with $\gamma$ being the nonlinear coefficient of the fiber. For anomalous dispersive fiber, this equation can be normalized by using the 'soliton' number $N$ defined as $N^2 = |L_D| / L_{NL}$ where $L_{NL}$ is the nonlinear length $L_{NL} = 1/\gamma |a_0|^2$.

$$i\frac{\partial \psi}{\partial \xi} = -\frac{1}{2}\frac{\partial^2 \psi}{\partial \eta^2} - N^2 |\psi|^2 \psi, \tag{11}$$

In the following, we will solve the NLSE using numerical simulations based on the split-step Fourier method [19].

We first consider the temporal evolution in an anomalous fiber for which the combination of dispersion and nonlinearity leads to a focusing behavior. The temporal intensity profile plotted for a fixed propagation distance $\xi = 0.5$ is shown on Fig. 3(a1) for three different light power level (expressed here through the soliton number). We can note several important points that stress how nonlinearity affects the temporal pattern. Whereas the positions of the oscillations that emerge on each side of the shadow are marginally affected, the amplitudes of these oscillations dramatically increase with increasing power and the oscillations become more and more abrupt. Such changes in the upper part of the signal have been the subject of recent discussions and can be interpreted in terms of solitons over finite background such as Akhmediev or Peregrine breathers [20, 21]. Similar structures linked to nonlinear Fresnel diffraction have also been observed in the spatial pattern resulting from the propagation in nonlinear Kerr material [22]. However, to the best of our knowledge, no study has reported the impact of nonlinearity on the central spot resulting from the constructive interference of the two edges. The level of this central Arago spot depends of the level of the signal, a higher amplitude of the input signal leading, in the anomalous regime of dispersion, to a decrease of the central Arago spot intensity (Fig. 3(a1)).

The case of a defocusing nonlinearity (i.e. a normally dispersive fiber) leads to a very different evolution as shown on panel (a2) of Fig. 3. In this case, the lateral oscillations that emerge from each edge are significantly reduced with increasing power. The central Arago spot intensity tends to become larger with increasing power. Ultimately, for higher propagation distances or power, the gap tends to be filled and dark solitonic structures may emerge as it has been shown in the temporal domain in the study of undular bores [23] and the interaction of a pair of delayed pulses [24] as well as in the spatial domain in defocusing materials [25, 26].

Both anomalous and normal regimes of propagation are summarized in Fig. 3 panel (b) showing that the evolution is monotonic with power, with an increase of the Arago central spot when $N$ increases in a defocusing medium (normal dispersion) and the contrary for the focusing medium (anomalous dispersion).

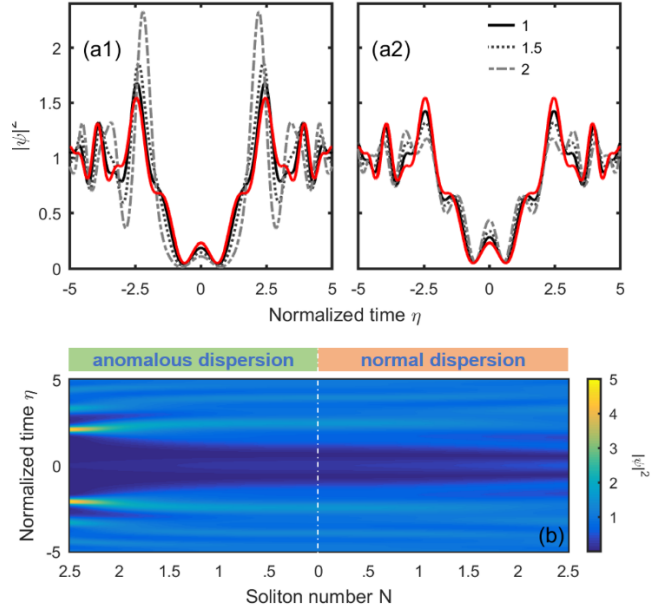

**Figure 3** – (a) Temporal intensity profiles at different input powers at a fixed propagation distance of $\xi = 0.5$. The result of linear propagation (red curves) is compared with the temporal profile obtained for a soliton number $N$ of 1, 1.5 and 2. Results achieved in the anomalous regime of dispersion (panel a1) are compared with the propagation in a normally dispersive fiber (panel a2). (b) Evolution of the temporal intensity profile at $\xi = 0.5$ with the initial soliton number $N$ in the anomalous and normal dispersion regimes.

## 3. Experimental setup

In order to confirm the trends predicted by the theoretical analysis and numerical simulations, we have implemented the experimental setup sketched in Fig. 4. The experiments rely on devices that are commercially available and used in the telecommunication industry. The initial temporal profile is obtained from a continuous wave laser at 1550 nm thanks to a Lithium Niobate intensity modulator operated at his point of maximum transmission and driven by an electrical pulse generator that delivers super-Gaussian pulses at a repetition rate of 2.5 GHz. A particular care has been devoted so that the resulting intensity profile presents a highly symmetric profile, with a strong platitude of the high level as well as a high extinction ratio to prevent parasitic interference between a residual unwanted background and the main structures [27]. The corresponding optical intensity profile is plotted on Fig. 4(b1) (blue solid line) and approximates the sharp edges of the ideal waveform (red solid line). The experimental intensity profile can be well fitted by an inverted second-order super-Gaussian pulse with a full-width at half duration of 40 ps, which is reasonably close from the ideal step edges. The initial optical spectrum recorded on a high resolution optical spectrum analyzer confirms the high level of symmetry of the pulse which is close to the Fourier limit. Let us note that for the study of the impact of nonlinearity, we have inserted a phase modulator that prevents the consequences of Brillouin backscattering.

The linear propagation experiment takes place in a set of optical single mode fibers with anomalous dispersion, i.e. smf-28 fibers [28]. The high level of dispersion ($\beta_2 = -20$ ps²/km) enables us to neglect the impact of second order dispersion. Using various spools with length between 0.5 and 5 km, we were able to record the longitudinal evolution of the field from 2.5 km of propagation up to 15 km. An alternate method to the use of such combination of fibers with a fixed length could be the use of a recirculating loop [29]. Note that chirped fiber Bragg grating could also be involved as a dispersive medium instead of a km-long fiber. The detection of the output pulse properties is ensured by a complex optical spectrum analyzer that enables us to get access both to the spectral intensity and phase profiles. It becomes therefore possible to measure the temporal phase and intensity profiles with a temporal resolution well below 1 ps. For this series of measurements, we have checked that the propagation was purely linear with no changes of the intensity spectral profiles.

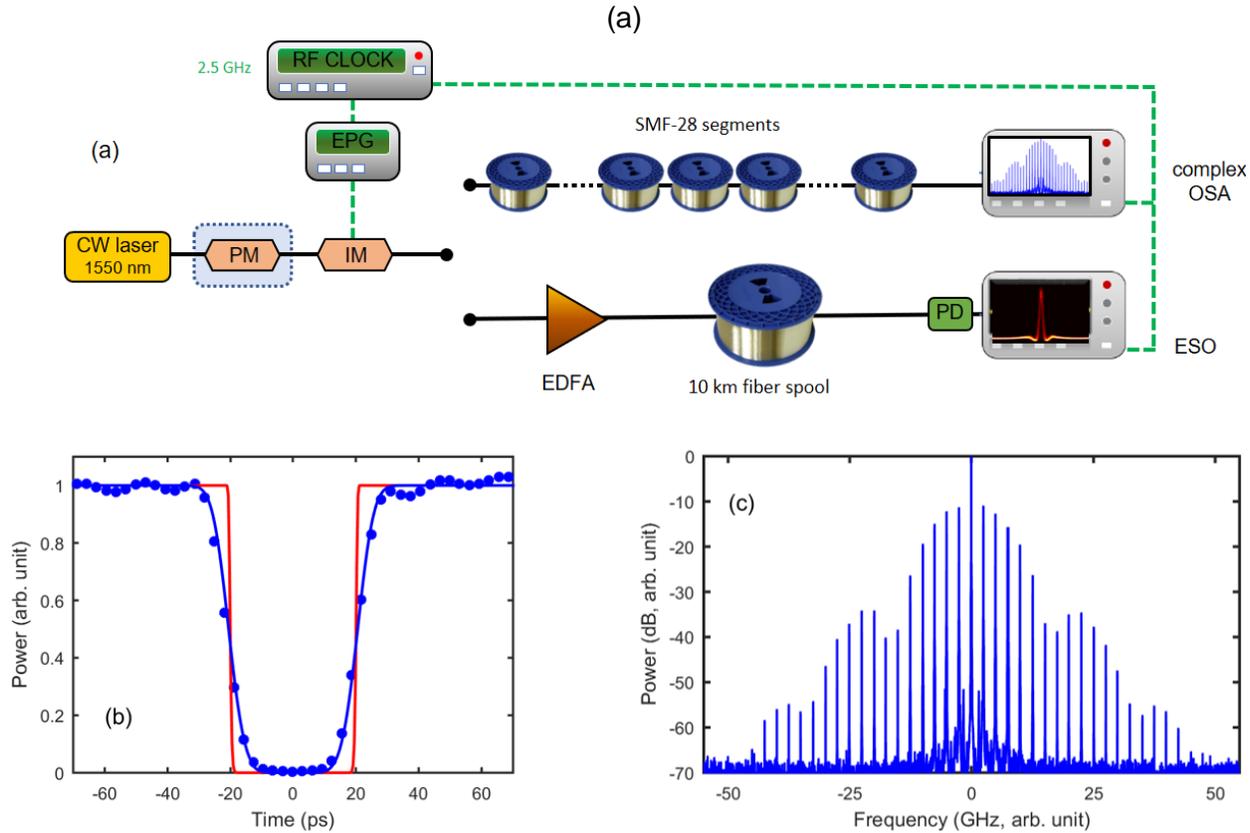

**Figure 4** – (a) Experimental setup. CW : continuous wave, PM : phase modulator, IM : intensity modulator, EPG : Electrical Pattern Generator, EDFA : Erbium Doped Fiber Amplifier, PD : Photodiode, ESO : Electrical Sampling Oscilloscope, OSA : Optical Spectrum Analyzer. (b) Temporal intensity profile obtained after modulation. The experimental results recorded with the complex spectrum analyzer (blue circles) are compared with a fit with a second-order super-Gaussian waveform (blue solid line). The red line represents the ideal inverted rectangular function. (c) Optical spectrum of the initial signal.

In order to carry out the study dealing with the nonlinearity, we have inserted an erbium doped fiber amplifier that delivers an output average power up to 23 dBm. Two types of fibers have been compared: (1) a SMF-28 fiber (identical to the previously used fiber, with a nonlinear coefficient $\gamma = 1.1$ $W^{-1}.km^{-1}$) with a length of 10 km and (2) a dispersion shifted fiber with a normal dispersion of -19 $ps^2$/km, a nonlinear coefficient of 2 $W^{-1}.km^{-1}$ and a length of 10.5 km. The two fibers therefore exhibit a rather similar level of integrated dispersion ($\approx 200$ $ps^2$ in absolute value) and in both cases, the low level of optical loss (0.2 dB/km) enables us to neglect the consequences of linear attenuation. The level of nonlinearity and input available powers allows to explore nonlinear reshaping up to $N \approx 2$. Given the additional phase modulation used to prevent Brillouin scattering, it was not possible to take advantage of the complex spectrum analyzer and to obtain the phase profile of the waveforms with increasing power. Therefore, in order to record the temporal intensity profiles, we used a photodiode with a high bandwidth (70 GHz) connected to a high-speed sampling oscilloscope with an electrical bandwidth exceeding 50 GHz.

## 4. Experimental results
### 4.1 Evolution in a purely dispersive medium

We first present the experimental results obtained in the linear regime of propagation. Details of three intensity and phase profiles are provided on Fig. 5(a) for three propagation distances (4, 8 and 12km). They are qualitatively in line with the analytical predictions made in section 2.2 and reported in Fig. 2(a). The dispersive propagation is marked by the emergence of oscillations on each edge. A central Arago spot progressively grows while the phase difference $|\Delta\varphi|$ between the central part and the surrounding plateau decreases.

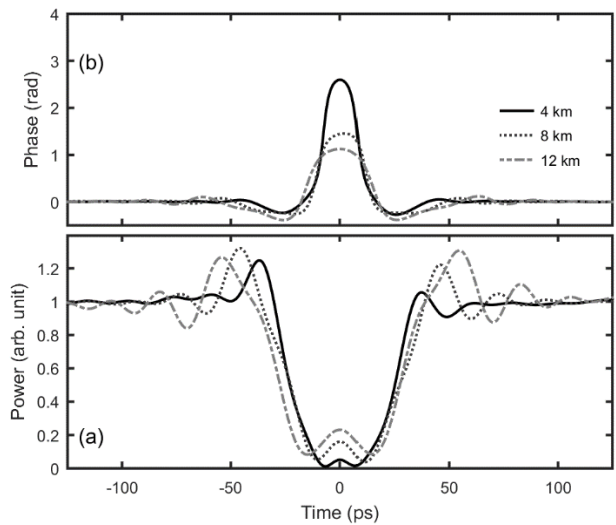

**Figure 5** – Experimental profiles of (a) the temporal intensity and (b) the temporal phase at three propagation distances (4, 6 and 8 km) corresponding to $\xi = .2, .4$ and $.6$ (solid black lines, doted grey lines and mixed grey lines respectively).

A more systematic study made every 500 m and presented in Fig. 6 enables us to reconstruct the longitudinal evolution of the temporal and phase intensity profiles (panels a) that are found in excellent agreement with the theoretical predictions. As propagation distance increases, the oscillations move away from the central part but their amplitude does not evolve significantly. On the contrary, the amplitude of the Arago central spot continuously increases and the phase is also affected. Panels b of Fig. 6 summarizes the quantitative evolution of the peak power in the central spot as well as the value of $|\Delta\varphi|$. These data are compared with the analytical predictions of Eq. (8) and (9) based on the assumption of an abrupt edge. Whereas the analytical trends are well reproduced by the experiment, some discrepancies are visible for short propagation distances. Indeed, whereas the analytics predicts a level of the Arago central spot that should be already detectable after 2 km, the experiment does not reveal the clear existence of the central spot. Moreover, at early stages of propagation, $|\Delta\varphi|$ predicted by the theory is higher than the experimental records. When taking into account the exact initial temporal intensity profiles and when using numerical simulations to solve Eq. (2), we can note that these discrepancies vanish. Using a second-order super-Gaussian pulse or perfect steps for the initial conditions leads to the same predicted result for large propagation distances (Fig.6 (b1-2) red and blue solid lines). We can note that experimentally, after 15 km of propagation, the central Arago spot is already a third of the power present on the plateau in excellent agreement with the theory.

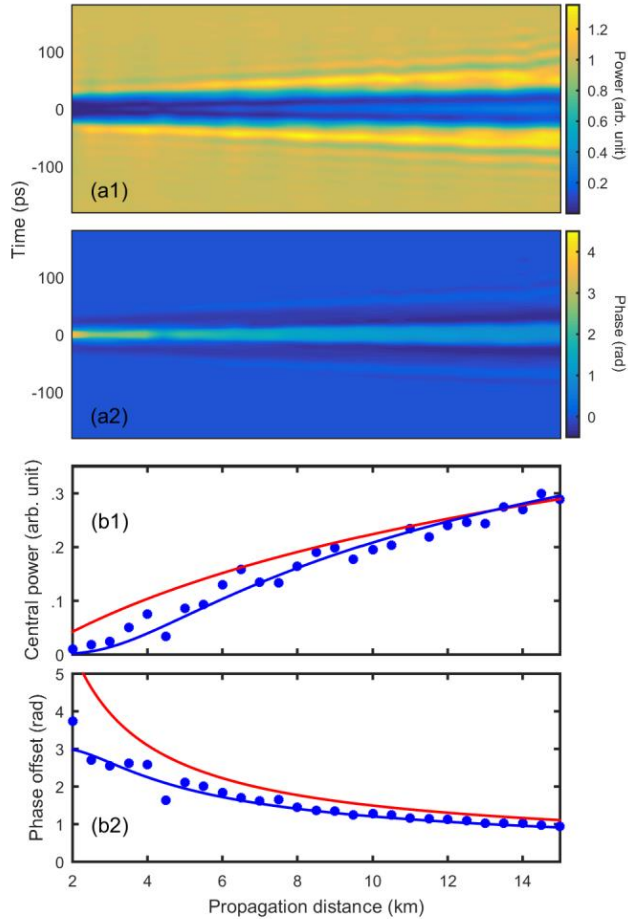

**Figure 6** – Experiment: longitudinal evolution of the temporal intensity and phase profiles (panels 1 and 2, respectively). The longitudinal evolution of the patterns are plotted on panels a. Panels b provides the details of the evolution for the peak power of the central spot and the phase offset $|\Delta\varphi|$. Experimental results (blue circles) are compared with analytical results provided by Eq. (8) and (9) (red curve) and results from numerical simulations taking into account a second-order super-Gaussian pulse as an initial condition (blue curve).

### 4.2 Evolution with nonlinearity

Our second set of experiments focused on the impact of the nonlinearity. The output intensity profiles obtained after propagation in 10-km long fibers with normal or anomalous dispersion are plotted for three levels of input power in Fig. 7(a). These results confirm the different trends that have been previously identified and discussed in section 2.3. For the focusing nonlinearity case (Kerr nonlinearity with anomalous dispersion), an increasing input power leads to the progressive development of the lateral oscillations. Note however that compared to numerical simulations of Fig. 3(a1), the level of the lateral peaks is decreased and their temporal duration increased as compared to theory due to the finite bandwidth of our electrical detection. Following the predicted trends, we observe a decrease of the central Arago spot when increasing the input power. The opposite behavior is observed for normally dispersive fibers: the central spot is increased whereas the lateral oscillations are lowered (see panel (a2) of Fig. 7).

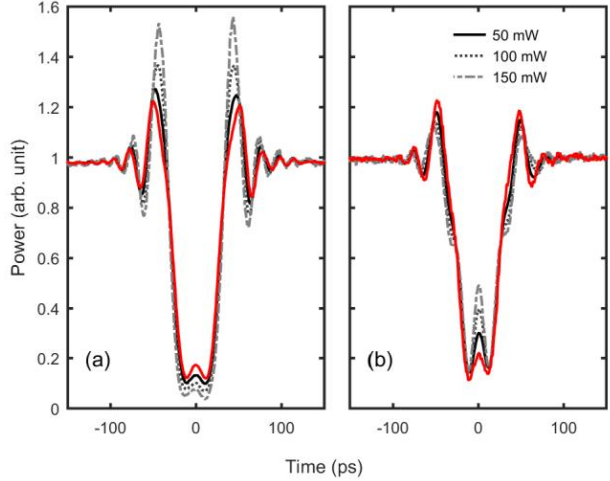

**Figure 7** – Experiment: temporal intensity profiles at different input powers at a fixed propagation distance of $\xi = 0.5$. The result of linear propagation (red curves) are compared with the temporal profile obtained for input average powers of 50, 100 and 150 mW (solid black lines, dotted grey lines and dashed grey lines respectively). Measurements performed in the anomalous (a) and normal (b) regime of dispersions.

A more systematic study of the output pattern according to the input power is reported on Fig. 8(a) and confirm those trends. In panel (b) of Fig. 8, we quantitatively summarized the evolution of the peak power of the Arago central spot according to the power and the regime of dispersion. The experimental results are in line with the results provided by the numerical integration of NLSE (red line) and are in excellent agreement with the numerical simulations taking into account the fiber loss and the initial second-order super-Gaussian temporal shape.

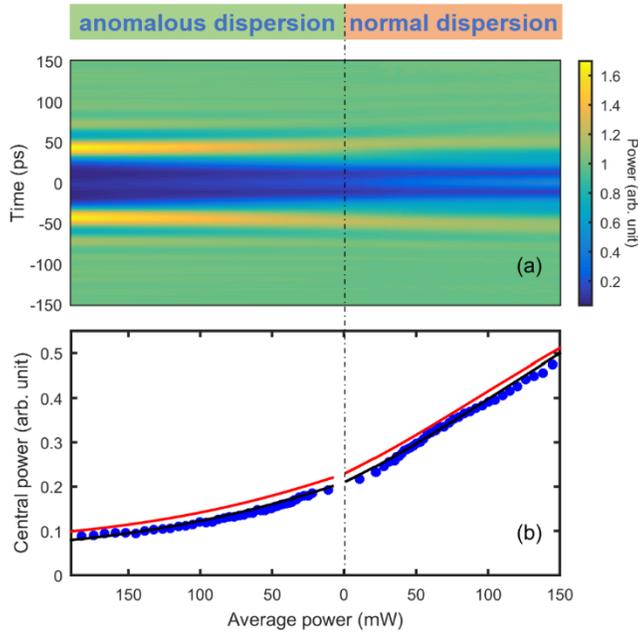

**Figure 8** – Experiment: evolution of the temporal intensity profile according to the initial input average power and according to the regime of dispersion. (a) Evolution of the output pattern. (b) Evolution of the peak power of the Arago central spot. Experimental results (blue circles) are compared with simulations based on the numerical integration of the NLSE with a perfect step edges as initial conditions (red curve) and with solving NLSE including losses and a second-order super-Gaussian pulse as the initial condition. (black curve).

## 5. Conclusion

In conclusion, we have investigated theoretically, numerically and experimentally the emergence of a bright spot in the initial shadow of a temporal signal. The studied case is the 1D temporal analog of the Arago spot formation that was firstly demonstrated by Fresnel to support his wave theory and acclaimed by Arago in 1817. We have showed that the spot progressively grows with the propagation distance whereas a phase difference exists between the spot center and the surrounding light. We have also demonstrated that Kerr nonlinearity affects the intensity of the Arago spot depending on the dispersion regime: a defocusing nonlinearity (Kerr effect and normal dispersion) increases the intensity of the spot whereas a focusing nonlinearity (Kerr effect and anomalous dispersion) tends to decrease it intensity with increasing power. The analytical and numerical results are fully confirmed by experiments conducted in optical fibers with normal and anomalous dispersion.

This work further demonstrates the space-time duality for the propagation of waves in the case of a simple 1D diffraction and has some relevance with the emergence of ghost pulses in optical communications [30]. Our analysis and experiments were conducted with continuous wave, but the results can be extended to the case of pulses with finite temporal extend. The space-time analogy can also be extended to other domains of physics where the ENSL is relevant such as Bose-Einstein condensates [31] or the propagation of oceanic waves [32]. We have here focused our discussion on the case the intensity profile exhibits strong intensity edges, but the discussion can be extended to the case of initial phase profiles or even frequency jumps [33, 34].

**Acknowledgement**

C.F. acknowledges the support by the Région Bourgogne Franche-Comté and the Institut Universitaire de France. H. R. acknowledges the support from the CNRS and Aix-Marseille University.